\documentstyle[12pt,amssymb]{article}

\begin{document}

\begin{titlepage}

\begin{center}

{\Large \bf Stationary Coverage of a Stochastic  \\ Adsorption-Desorption
Process \\  with Diffusional Relaxation \\ }

\vspace{1.5cm}

{\large J. Ricardo G. de Mendon\c{c}a\footnote{Departamento de F\'{\i}sica,
Universidade Federal de S\~{a}o Carlos, 13565-905, S\~ao Carlos, SP, Brazil.}
and M\'{a}rio J. de Oliveira\footnote{Instituto de F\'{\i}sica, Universidade 
de S\~{a}o Paulo, 05315-970, S\~{a}o Paulo, SP, Brazil.}}

\vspace{1.5cm}

{\large \bf Abstract}

\vspace{0.5cm}

\parbox{12.0cm}{We show that it is possible to derive the stationary
coverage of an adsorption-desorption process of dimers with
diffusional relaxation with a very simple ansatz for the stationary
distribution of the process supplemented by a hypothesis of global
balance. Our approach is contrasted to the exact result and we seek
to understand its validity within an instance of the model.}

\vspace{1.5cm}

Key words: stochastic lattice gas; dimers; stationary state; free-fermions.

\end{center}

\end{titlepage}

\section{Introduction}

The use of master equations is one of the most promising techniques
in the study of nonequilibrium statistical systems, their success
stemming both from their nice mathematical properties and from their
phenomenological character, making them a modelling tool in a great
variety of situations.

In order to set up the master equation for a model system, one is
asked to give the transition rates between the different possible
configurations of the system. Under rather weak conditions, namely
that each configuration must be reachable from every other, one can
assert that as time tends to infinity the probability distribution of
configurations will tend, for finite systems, to a unique nonzero
stationary distribution. In the case detailed balance holds among the
rates, a condition that can be easily verified by perusal of the
Kolmogoroff's criterion$^{(1)}$, the matrix of the transition rates
can be cast in the form $DSD^{-1}$ with $D$ diagonal and $S$
symmetric$^{(2)}$, and the system's stationary distribution will be
an equilibrium one such that we can assign to it an effective energy
$H=\ln P_{s}$ simply related to the rates. When detailed balance
does not hold, on the contrary, the determination of the stationary
distribution becomes considerably a more involved task, its structure
being in general almost completely unknown. Albeit such a
distribution will exist, and a number of techniques have developed
aiming at the evaluation, at least in an approximate fashion, of an
energy function also in these cases; noteworthy among these is the
correlation method$^{(3,4)}$.

The purpose of this paper is to show how a very simple ansatz for the
stationary distribution plus a hypothesis of global balance can be
used to clarify some results for a reaction-diffusion process that
was in vogue some time ago$^{(5-7)}$. The process consists of
adsorption and desorption of dimers together with asymmetric
diffusion of monomers on a one-dimensional lattice. The previously
derived results for the process were obtained using a
Schr\"odinger-like description of the master equation that renders
the infinitesimal generator of the Markov semigroup a magnetic
Hamiltonian aspect, thus allowing for techniques first developed in a
magnetic context to be used in the stochastic context with equal profit.

The paper goes as follows. In Section 2 we derive the
Schr\"odinger-like form of the master equation in the special case
where only two states per site are present; we believe the derivation
given there is clearer than the ones usually encountered, though it
is not properly a novelty. We also define the model and state the
results we are interest in. In Section 3 we make our contribution to
the understanding of the stationary coverage of the model and in
Section 4 we discuss our result in the light of previous exact
results obtained for this quantity.

\section{The Time Evolution Operator}

Let us begin with the master equation. We consider a one-dimensional
lattice $\Lambda \subseteq {\Bbb Z}$ of $|\Lambda|=L$ sites, and
attach to each site $\ell \in \Lambda$ a random variable
$\sigma_{\ell}$ taking values on $\omega = \{-1,+1\}$, the state
space of the whole lattice being given by $\Omega =
\omega^{\Lambda}$. If $W(\sigma^{\prime},\sigma)$ denotes the rate of
transition between the configurations $\sigma$ and $\sigma^{\prime}$
($\sigma \to \sigma^{\prime}$), $\sigma, \sigma^{\prime} \in \Omega$,
and $P(\sigma,t)$ is the probability of realization of a particular
configuration $\sigma$ at instant $t$, we write the master equation as
\begin{equation}
\frac{\partial P(\sigma,t)}{\partial t}=\sum_{\sigma^{\prime} \in \Omega}
\left[ W(\sigma,\sigma^{\prime})P(\sigma^{\prime},t)-
W(\sigma^{\prime},\sigma)P(\sigma,t) \right].
\end{equation}
Using the fact that in a one-dimensional lattice of two-state
variables $|\Omega|=2^L= |\{R\subseteq\Lambda\}|$ and also that the
only possible channel of collision for a $\sigma_{\ell}$ is
$\sigma_{\ell} \to \sigma_{\ell}^{\prime}=-\sigma_{\ell}$, we write
the transition rates most generally as
\begin{equation}
W(\sigma,\sigma^{\prime})=\sum_{R\subseteq \Lambda}W_R(\sigma^{\prime})
\delta(\sigma^R-\sigma^{\prime}),
\end{equation}
where by $\sigma^R$ we mean the configuration which
equals $\sigma$ except for the sites in the region $R$, where
$\sigma_{r}^{R}=-\sigma_r$, $r\in R$, and the delta is a product of
Kronecker's deltas over the whole lattice. We then rewrite Eq.\ (1) as 
\begin{equation} \frac{\partial P(\sigma,t)}{\partial t}=
\sum_{R\subseteq \Lambda}\left[ W_R(\sigma^R)P(\sigma^R,t)-
W_R(\sigma)P(\sigma,t)\right].
\end{equation}

We now explicitly introduce linear vector spaces in the description
of the structure of Eq.\ (3). To do this we turn $\omega=\{-1,+1\}$
into $\omega={\Bbb C}^2$ and $\sigma$ into $|\sigma\rangle =
\bigotimes_{\ell \in \Lambda}|\sigma_{\ell}\rangle$, the state space
now being given by $\Omega=\bigotimes_{\ell \in \Lambda}\omega$.
Taking an orthonormal basis $\{|\sigma\rangle\}$ for $\Omega$ we
write 
\begin{equation}
|P(t)\rangle =\sum_{\sigma\in \Omega}P(\sigma,t)|\sigma\rangle 
\end{equation}
for the generating vector of the probability densities 
$P(\sigma,t)=\langle\sigma|P(t)\rangle$. 
In the space of the linear operators acting on $\Omega$
we define $\hat{X}_R$ and $\hat{W}_R$ by their actions
\begin{equation}
\hat{X}_R |\sigma\rangle = |\sigma^R\rangle \hspace{0.5cm}
{\rm and}\hspace{0.5cm} \hat{W}_R |\sigma\rangle =
W_R(\sigma)|\sigma\rangle .
\end{equation} 
Multiplying Eq.\ (3) by $|\sigma\rangle$ and summing over $\Omega$, 
with the help of the above defined operators we eventually arrive at
\begin{equation}
\frac{\partial |P(t)\rangle}{\partial t}=-\hat{H}|P(t)\rangle ,
\end{equation}
with 
\begin{equation}
\hat{H}=\sum_{R\subseteq\Lambda}(\hat{1}-\hat{X}_R)\hat{W}_R
\end{equation}
the infinitesimal generator of the Markov semigroup
$\hat{T}(t) = \exp (-\hat{H}t)$.  Eq.\ (6) is in the desired
Schr\"odinger-like form. If we use for the matrices of the
$\hat{X}_R$ and $\hat{W}_R$ operators the basis of Pauli matrices
with $\sigma^z$ diagonal we readly see that 
\begin{equation}
\hat{X}_R = \prod_{r\in R}\sigma_{r}^{x} \hspace{0.5cm} 
{\rm and} \hspace{0.5cm} \hat{W}_R = W_R(\sigma^z) , 
\end{equation} 
where $\sigma_r^{\alpha}$ stands for $\sigma_r^{\alpha}=\hat{1}\otimes
\ldots \otimes \hat{1} \otimes \sigma^{\alpha} \otimes \hat{1}
\otimes \ldots \otimes \hat{1}$, the $\sigma^{\alpha}$ being in the
$r$-th position, $\alpha = x,y,z$.

The process we are interested in is one in which pairs of particles
adsorb with rate $\epsilon$ and desorb with rate $\epsilon^{\prime}$
from the lattice and which also admits diffusion of monomers to the
right with rate $h$ and to the left with rate $h^{\prime}$
$^{(5-7)}$.  Identifying the presence of a particle in the site
$\ell$ with the eigenvalue $+1$ of $\sigma^z_{\ell}$ we obtain from
Eqs.\ (7) and (8) the two-body evolution operator 
\begin{eqnarray}
\hat{H}_{\ell,\ell+1} & = & (1-\sigma^x_{\ell}\sigma^x_{\ell+1})
\frac{1}{4}\left[ \epsilon (1-\sigma^z_{\ell})(1-\sigma^z_{\ell+1})+
\epsilon^{\prime}(1+\sigma^z_{\ell})(1+\sigma^z_{\ell+1}) \right. \nonumber \\
& + & \left. h(1+\sigma^z_{\ell})(1-\sigma^z_{\ell+1}) +
h^{\prime}(1-\sigma^z_{\ell})(1+\sigma^z_{\ell+1})\right] ;
\end{eqnarray} 
notice that the collision operator $\hat{X}_{\ell,\ell+1} = 
\sigma^x_{\ell}\sigma^x_{\ell+1}$ is common to all the
elementary processes of the model. The total evolution operator can
be obtained summing the two-body operator over the sites of the
lattice provided some boundary condition is given.

The operator in Eq.\ (9) is a very interesting and very complicated
one. From the magnetic point of view it is an $XXZ$ Heisenberg
ferromagnet with both $XY$ and Dzyaloshinskii-Moriya (DM) in-plane
interactions with pure imaginary couplings, plus an external field
and, for open boundary conditions, a surface term. We can see,
developing the products in (9), that the coupling in the $XY$ term is
associated to the difference of the adsorption and desorption rates,
while that in the DM term is associated to the asymmetry in the
diffusion. It is interesting to note that an asymmetric diffusion
breaks the chiral symmetry that the system would have otherwise, and
that this reflects in the DM term $( \vec{\sigma}_{\ell}\times
\vec{\sigma}_{\ell+1})\cdot \hat{e}_z $, that just does exactly the
same thing.

With periodic boundary conditions the total operator resulting from
Eq.\ (9) have been exactly solved for a number of choices of the
rates$^{(5-7)}$.  For example, for $\epsilon = \epsilon^{\prime} = h
= h^{\prime}$ it reduces to the Ising model diagonal in the
$\sigma^x$ representation, and for $\epsilon = \epsilon^{\prime} \neq
h = h^{\prime}$ it can be cast, after a rotation over the bipartite
lattice, to a ferromagnetic $XXZ$ Hamiltonian in one of its massive
phases. In particular, for $\epsilon + \epsilon^{\prime} = h +
h^{\prime}$ the evolution operator turns out to be quadratic in
$\sigma^{\pm}=\frac{1}{2}(\sigma^x \pm \dot{\imath}\sigma^y)$ and
thus can be solved in terms of free fermions$^{(5,6)}$. It is to this
instance of the model that we want to adress our observations.

\section{The Stationary Coverage} 

It is possible to interpret domain walls in the Glauber kinetic Ising
model$^{(8)}$ as particles in a reaction-diffusion scenario$^{(9)}$.
Although in the original model domain walls diffuse symmetricaly,
this can be relaxed to allow for biased diffusion without departing
too much from the original structure of the equations of the
model$^{(5)}$, though detailed balance and the connection with the
equilibrium distribution of the Ising model do not hold in this case
anymore. On the other hand the relationship between the Glauber model
and free fermions have been established already a long time
ago$^{(10)}$. All this led to the conclusion that whenever
$\epsilon+\epsilon^{\prime}=h+h^{\prime}$ in the model described in
the last section, one can associate to it a free fermion evolution
operator that is also the evolution operator of an asymmetric version
of the Glauber model.

The above mentioned relantionships have made it possible to compute
the time-dependent density profile of the model in the free fermion
case, with the result that the stationary coverage is given
by$^{(5,6)}$ 
\begin{equation} \lim_{t \to \infty} \varrho(t) = \varrho_s =
\frac{1}{1+\sqrt{\epsilon^{\prime}/\epsilon}}.
\end{equation}
We would like to rederive this result without recourse
to the free fermion constraint.

Guided by the duality between domain walls and particles and by the
product form of the equilibrium distribution of the Glauber model
(since $\sigma_{\ell}^z \sigma_{\ell +1}^z \to \tau_{\ell}^z$ under
the duality transformation) we ask whether we can derive any results
by postulating a stationary distribution of the form 
\begin{equation}
P_{s}(\sigma) = \prod_{\ell \in \Lambda}P_{s}(\sigma_{\ell}) =
\frac{1}{Z}\exp \left( J\sum_{\ell \in \Lambda}\sigma_{\ell} \right) ,
\end{equation}
where $Z=(2\cosh J)^L$ is a normalizing constant. In
doing this we were particularly inspired by the results in Ref.\ 4.
From the master equation (3) we see that $P_s(\sigma)$ will be
stationary if 
\begin{equation} \sum_{R \subseteq \Lambda}\left[W_R(\sigma^R)P_s(\sigma^R)-
W_R(\sigma)P_s(\sigma)\right]=0 ,
\end{equation}
which is not the condition of detailed balance, since
we are not requiring each term of the sum to vanish, but only the
whole sum to vanish instead. Looking at the rate operator $\hat{W}_R$
in Eq.\ (9) we see that we can write it as 
\begin{equation}
\hat{W}_{\ell, \ell+1}= A+B\sigma_{\ell}^z+C\sigma_{\ell+1}^z+
			D\sigma_{\ell}^z\sigma_{\ell +1}^z
\end{equation}
with $A=\frac{1}{4}(\epsilon^{\prime} + \epsilon + h + h^{\prime}),
\  B=\frac{1}{4}(\epsilon^{\prime} - \epsilon + h - h^{\prime}),
\  C=\frac{1}{4}(\epsilon^{\prime} - \epsilon - h + h^{\prime})$
and $D=\frac{1}{4}(\epsilon^{\prime} + \epsilon - h - h^{\prime})$;
notice that under the free fermion condition $D=0$,
corresponding to the vanishing of the ``many-body'' term in Eq.\ (9).
Since we are dealing with a basis diagonal in the $\sigma_{\ell}^z$'s
we will treat them in Eq.\ (13) as if they were c-numbers. From
Eqs.\ (12) and (13) it follows that $P_s(\sigma)$ will be stationary if
\begin{equation}
\sum_{\ell \in \Lambda}\left\{W_{\ell,\ell+1}(\sigma^{\ell, \ell+1})
\exp \left[ -2J(\sigma_{\ell}+\sigma_{\ell+1})\right] - 
W_{\ell,\ell+1}(\sigma)\right\} = 0,
\end{equation}
and taking into account the translational invariance
of the system we arrive at the following condition on the coupling
constant $J$, 
\begin{equation} 
\tanh 2J = -\frac{B+C}{A+D} = \frac{\epsilon - \epsilon^{\prime}}
{\epsilon + \epsilon^{\prime}} ,
\end{equation} 
that is, 
\[
e^{-4J} = \frac{\epsilon^{\prime}}{\epsilon}, 
\]
and the stationary coverage per site reads 
\begin{equation} \varrho_s = \frac{e^J}{e^J + e^{-J}} = 
\frac{1}{1+\sqrt{\epsilon^{\prime}/\epsilon}} ,
\end{equation}
as in Eq.\ (10).

The above result is rather surprising. It states that in the
stationary regime mean-field analysis, as expressed by (11), holds
exact. It also tells us that despite the fact that when $h\neq
h^{\prime}$ a current of particles establishes in the substrate, in
the stationary state the concentration of particles only depends upon
the adsorption-desorption rates. Similar results have been obtained
for the case $h=h^{\prime}=0$, but in this case the model shows
microscopic reversibility and a ``Boltzmann weight analysis'' can be
performed$^{(11)}$, which in the case being is not an a priori valid
procedure.

\section{Conclusions} 

The expressions for the concentration of particles in the stationary
state given by Eqs.\ (10) and (16), though identical, were obtained by
very different techniques. Our result is mean-field-like but anyway we were
able, with the aid of a hypothesis concerning global balance, 
to reproduce the exact result.

It is well known that both the symmetric and asymmetric simple
exclusion processes have product form stationary distributions like
that in \mbox{Eq.\ (11)}, i.e., they both have white noise as
invariant measures, as is sometimes said. This however should not
lessen one's surprise in having $P_s(\sigma)$ as in Eq.\ (11) for the
process we are dealing with in this paper since it has, besides one
particle exclusion, adsorption of dimers in which two adsorbed
particles together prevent further adsorption, and when all four
rates that define the model are non-null it is not possible by any
amount of sublattice-mapping funambulism to turn the model into an
instance of a simple exclusion process. Anyway, for $\epsilon =
\epsilon^{\prime}=0$ our analysis breaks down, for in this case
Eq.\ (15) becomes ill-defined. On the other hand, when
$h=h^{\prime}=0$ but $\epsilon \neq 0 \neq \epsilon^{\prime}$ the
same result for $\varrho_s$ holds$^{(5,11)}$, and it thus appears
that a finite amount of adsorption-desorption contrives $\varrho_s$
to be given by Eq.\ (16), wich is indeed a curious result.

One possible explanation for this is that due to the similitude of
the model with the Glauber model through a site-bond transformation
one expects that in the particle scenario the expression given by
Eq.\ (11) is a reasonable guess. It should be mentioned however that
the exact duality demands $h=h^{\prime}$ which in our approach is not
necessary.

One can naively say that the stationary coverage does not depend on
the asymmetry of diffusion due to the periodic boundary conditions:
one then simply performs a Galilei transformation $\ell \to \ell
+(h-h^{\prime})t$ on the whole lattice and follows the time
dependencies on the stationary state on the new reference frame,
where the diffusion happens to be symmetrical and the exact duality
with the Glauber model is recovered. This procedure, if plausible in
some special circumstances, to date has been proved right only for a
special class of initial distributions, the completely random ones,
and in the free fermion point 
$\epsilon+\epsilon^{\prime}=h+h^{\prime}$ $^{(7)}$.

We would like to stress that we {\it guessed\/} the product form
stationary distribution Eq.\ (11), and it happened that it contained
coupling constants enough to allow for the derivation of Eq.\ (16),
in the case only one coupling constant $J=J(\epsilon,
\epsilon^{\prime})$. It could equally well have happened that other
couplings, say $K\sigma_{\ell}\sigma_{\ell +1}$ with
$K=K(\epsilon,\epsilon^{\prime},h,h^{\prime})$, would have been
necessary to derive useful results.

As a final remark we notice that the hypothesis of global balance and
its connection with periodic boundary conditions can be further
exploited, e.~g.\   using open boundaries and adding the necessary
currents of particles at the ends so to satisfy the same global
balance equations, with $P_s(\sigma)$ possibly modified to account
for the open boundaries.

\section*{Aknowledgments} 

The authors would like to aknowledge the Brazilian agencies CNPq and FAPESP
for financial support.

\end{document}